\documentstyle[prl,aps,twocolumn,psfig,rotate,amsfonts,amssymb]{revtex}
\begin{document}
\def\u{\bbox}
\def\mathcal#1{{\cal #1}}
\def\mat#1{{\it #1}}
\def\d{\displaystyle}
\def\goldenmean{\gamma}
\draft
\pagestyle{myheadings}
\markright{Phys. Rev. Lett. {\bf 79}, 3881 (1997). }


\title{Kolmogorov-Arnold-Moser--Renormalization-Group Analysis of Stability 
in Hamiltonian Flows}

\author{M. Govin, C. Chandre, and H. R. Jauslin}

\address{Laboratoire de Physique, CNRS, Universit\'e de Bourgogne,
BP 400, F-21011 Dijon, France}
\maketitle
\begin{abstract}
We study the stability and breakup of invariant tori in Hamiltonian flows
using a combination of Kolmogorov-Arnold-Moser (KAM) theory and 
renormalization-group techniques.
We implement the scheme numerically for a family of
Hamiltonians quadratic in the actions to analyze the strong coupling 
regime. 
 We show that the KAM iteration converges up to the critical coupling
at which the torus breaks up. Adding a renormalization consisting of 
 a rescaling of phase space and a shift of
resonances allows us to determine the critical coupling with higher accuracy.
 We determine a nontrivial fixed point and its universality properties.
\end{abstract}

Regular Hamiltonian dynamics is characterized by the existence of as many
independent conserved quantities as degrees of freedom $d$. As a
consequence, each trajectory is confined to evolve on an invariant torus
of dimension $d$. This global regularity is destroyed in general by any 
small perturbation: some of the invariant tori break up and chaotic
trajectories appear. However, many invariant tori survive and are just
slightly deformed. The persistence of these tori has strong implications
on the global phase space structure of the dynamics. They strongly limit
the spreading of chaos and prevent global ergodic behavior. In particular,
for two degrees of freedom, they block completely the long distance
diffusion in phase space. In this context, it is important to understand
how invariant tori are destroyed as the size of the perturbation is 
increased.

The Kolmogorov-Arnold-Moser (KAM) technique was developed originally 
to prove the stability 
under small perturbations of a large fraction of the invariant tori
of an integrable Hamiltonian. 
It has been later applied to a very large spectrum of phenomena both in
 classical and in quantum mechanics. The KAM 
theorem\cite{kolmogorov,arnold,moser} states that, if the frequency satisfies a
diophantine condition, and the size of the perturbation $\varepsilon$ is smaller
than an $\varepsilon_0$, then a torus of that frequency will be stable.
The proof is based on an iterative algorithm to construct the invariant 
tori. Each step of the KAM iteration consists of a coordinate
transformation that reduces the size of the perturbation from order
$\varepsilon$ to $\varepsilon^2$.

However, it was remarked early that the value $\varepsilon_0$ obtained
in the proofs\cite{cellettichierchia}, for which the iteration is shown
to converge, is much smaller than the critical
coupling $\varepsilon_c$ at which the torus is known\cite{mackaypercival}
 to become unstable and to break up into a Cantor set  
(Aubry-Mather set)\cite{aubry,mather}.
Since there is no  physical
phenomenon associated with $\varepsilon_0$, the  difficulty to prove the
 convergence
of the KAM iteration for $\varepsilon\in[\varepsilon_0,\varepsilon_c)$
was interpreted as a shortcoming of the iterative approach.
A first goal of the present Letter is to provide numerical evidence that 
the KAM iteration can be expected to converge all the way to the critical
coupling $\varepsilon_c$. 

Renormalization-group (RG) ideas were proposed for the analysis of the 
breakup of
invariant tori\cite{escandedoveil,escande,gallavotti,mackay}
based on the observation of self-similar scaling properties
\cite{kadanoff,shenker,rand}, and inspired on the renormalization
theory of phase transitions\cite{wilson}.
 For Hamiltonian
flows, Escande and Doveil set up an approximate renormalization
scheme\cite{escandedoveil,escande} that consisted of a
combination of KAM transformations with a rescaling and shift of resonances.
 The 
approximation consisted in retaining at  each step only two Fourier modes,
corresponding to the main resonances, and neglecting all the other modes
as well
as all terms of order higher than two in the momentum. Although the
approximation seems very drastic, since the effect of the neglected terms  
could not be controlled, the general ideas of the renormalization
group approaches suggest that if the essential features are kept
in the approximate
scheme, the neglected parts  may turn out to be {\it irrelevant}. This can of
course only be firmly justified by treating the problem without the approximation, 
and establishing which degrees of freedom are relevant and which ones are
irrelevant.

An outstanding question in this context has been  whether this approximate
renormalization scheme can be improved in a systematic way, in order to 
understand the properties of an exact renormalization scheme. 
The idea of the renormalization approach is to iterate a transformation in the
space of Hamiltonians. For  couplings $\varepsilon<\varepsilon_c$ the iteration 
should converge to a {\it trivial fixed point}. This can be considered as an
alternative version of the KAM theorem\cite{koch}. 
The set of Hamiltonians with critical coupling $\varepsilon_c$ form a surface
in the space of Hamiltonians. The renormalization transformation maps this surface
into itself. The  main hypothesis of the
renormalization group approach
is that there should be another  {\it nontrivial fixed point} on the critical
surface that is attractive for  Hamiltonians on that surface. Its existence has
strong  implications concerning universal properties in the mechanism of
the breakup of invariant tori. 

The analysis of the approximate scheme of Escande and
Doveil\cite{escandedoveil,escande}, and conceptually similar 
studies on the dynamics of  area-preserving maps\cite{mackay},
 give support  to the
validity of the general picture. In order to give a firm basis to this approach, 
some essential points need to be established:
since the renormalization transformation contains at each step  a
KAM transformation, it is indispensable to rely on the existence of a well
defined KAM transformation for $\varepsilon$ all the way to $\varepsilon_c$. 
Furthermore the KAM iteration should either converge for all
 $\varepsilon<\varepsilon_c$
or at least it should do so when combined with the rescaling
and resonance shift.

Our approach has been to implement the KAM algorithm numerically  for
Hamiltonians with two degrees of freedom, quadratic in the action
variables ${\u A}=(A_1,A_2)$, of the form
\begin{equation}
\label{hamiltonian}
H({\u A},{\u \varphi})=
\frac{1}{2}m({\u \varphi})( {\u \Omega}\cdot{\u A})^{2}
+\lbrack {\u \omega}_{0}
+\varepsilon g({\u \varphi}){\u \Omega} \rbrack \cdot{\u A}
+\varepsilon f({\u \varphi}) \ ,
\end{equation}
where ${\u \varphi}=(\varphi_1,\varphi_2)$ are two angles, ${\u \omega_0}$ is the 
frequency vector of the considered torus, and ${\u \Omega}$ is some other constant
vector.
This type of model can describe, e.g., a charged particle interacting with two waves
in a plasma\cite{escande}, and  has a wide range of applications.

First we study the convergence properties of the  KAM iteration. Then 
we define a renormalization transformation (KAM-RG) by combining it with a shift 
of resonances, and a rescaling of momentum and energy.

We show that the KAM algorithm can be used to determine the critical coupling
$\varepsilon_c$. We then show that this can also be achieved with 
the KAM-RG transformation, and that it is much more efficient, in that very
high precision can be obtained already with few Fourier coefficients.

We establish the existence of a nontrivial fixed point by iterating the KAM-RG
transformation on the critical surface. We characterize the nontrivial fixed
 point Hamiltonian by its Fourier representation with a very large number
of coefficients (around 4000). 

We calculate the critical exponents by linearizing the renormalization
map around the fixed point. The result shows that there is only one 
unstable direction, transverse to the critical surface.

\paragraph*{KAM iteration.---}
We consider a Hamiltonian  of the form (\ref{hamiltonian}), where 
$\varepsilon g({\u \varphi}){\u \Omega}$ and
$\varepsilon f({\u \varphi})$ are the perturbations.
The KAM procedure consists of eliminating the perturbation 
of order $\varepsilon$ by a canonical transformation, which in turn produces terms
of order $\varepsilon^2$. In the approach we take, following W. Thirring
\cite{thirring,commentaire}, $m({\u \varphi})$ is of order one and does not need to be
eliminated. The point is that we want to establish the existence of a torus
of frequency ${\u \omega}_0$. If $(\ref{hamiltonian})$ can be transformed
into a Hamiltonian of the same form but with $f=0$, $g=0$, with arbitrary 
$m({\u \varphi})$,  the equations of motion show immediately
that at ${\u A}=0$ there is  an invariant torus
of frequency ${\u \omega}_0$. The great advantage of this approach is that 
the KAM transformation can be taken such that the transformed  Hamiltonian 
is also quadratic in the actions, i.e. the KAM transformation is a mapping among 
Hamiltonians of the form $(\ref{hamiltonian})$ (${\u \Omega}$ is not changed).
This is very convenient for a numerical implementation, since at each step of the
iteration the Hamiltonian is  completely determined by three functions $f,g,m$ of
the angles ${\u \varphi}$, that can be represented by their Fourier coefficients. Thus
the only approximation involved is the representation of the functions  by a finite
number of coefficients.  In our implementation we take all the Fourier modes
with wavenumber ${\u k}\in {\Bbb Z}^2$ in the square 
$\mathcal{C}_L$ of length $2L+1$ centered at $(0,0)$.

We perform the KAM transformation as a canonical change of coordinates 
$({\u A},{\u \varphi}) \mapsto
({\u A}',{\u \varphi}')$ in the Lie representation\cite{deprit}:
\begin{eqnarray}
H'({\u A}',{\u \varphi}')&=&
e^{+\varepsilon\hat{S}({ A},{ \varphi})}
H({\u A},{\u \varphi})\mid_{({ A}',{ \varphi}')}\nonumber\\
&=&
H+\varepsilon\lbrace S,H\rbrace
+\frac{\varepsilon^{2}}{2!}\lbrace S,\lbrace S,H\rbrace\rbrace\ldots
\end{eqnarray}
generated by a function $S$, linear in the actions ${\u A}$,  of the form
\begin{equation}
S({\u A},{\u \varphi})=
Y({\u \varphi})\, {\u \Omega}\cdot{\u A}
+Z({\u \varphi})+a\, {\u \Omega}\cdot{\u \varphi}\ .
\end{equation}
The functions $Y({\bbox \varphi})$, $Z({\u \varphi})$ and the
constant $a$ are chosen such that the terms of order $\varepsilon$ cancel out.
We start with the same initial Hamiltonian as in
Refs.\cite{cellettichierchia,escande}: 
\begin{equation}
\label{hamiltonieninit}
H({\u A},{\u \varphi})=\frac{1}{2}({\u \Omega}\cdot{\u A})^{2}
+{\u \omega}_{0}\cdot{\u A}+\varepsilon f({\u \varphi}) \ ,
\end{equation}
where ${\u \Omega}=(1,0)$, ${\u \omega}_0
=(1/\goldenmean,-1)$,
$\goldenmean=(1+\sqrt{5})/2$, and
$f({\u \varphi})=\cos({\u \nu}_1\cdot{\u \varphi})
+ \cos({\u \nu}_2\cdot{\u \varphi})$, where 
${\u \nu}_1=(1,0)$ and ${\u \nu}_2=(1,1)$.
We perform an iteration of the above KAM transformation, representing all the
functions by their Fourier series truncated by retaining only the coefficients in
the square $\mathcal{C}_L$. For fixed $L$ we take successively larger couplings
$\varepsilon$ and determine whether the iteration converges to a Hamiltonian with 
$f=0,\,  g=0$ or diverges. By a bisection procedure we determine the breakup
coupling $\varepsilon_c(L)$. We repeat the calculation with larger numbers
of Fourier coefficients, to obtain a more accurate approximation. In Fig.\ 1
we show $\varepsilon_c(L)$ i.e.  the dependence of the breakup coupling on the
number of Fourier coefficients retained. We observe that $\varepsilon_c(L)$ 
decreases with $L$ in a stepwise manner. It stays essentially constant except at the
points where a new rational approximant of the frequency ${\u \omega_0}$ is included,
corresponding to a resonance at the next smaller scale. The size of the jumps
diminishes approximately geometrically, and we can extrapolate to obtain the value
$\varepsilon_c(L) \to 0.0276$. This value is close to the
breakup $\varepsilon_c=0.0275856$ obtained by the Greene
criterion\cite{cellettichierchia,greene} which is surmised to yield the exact
value.

The calculation provides an explanation of the fact
that this global convergence is not easy to establish analytically. 
The KAM theorem is proven by showing that for $\varepsilon<\varepsilon_0$
the iteration is contractive. We observe in the numerical implementation of the
iteration that for larger $\varepsilon$ the convergence is not monotonic: in the
first KAM steps the norm of the perturbation can actually grow and oscillate, until it
reaches a region where it is contractive.

\paragraph*{KAM-renormalization group transformation.---}
In this section we show that by combining the KAM transformation with a
rescaling and a shift of resonances we can greatly improve the determination
of the breakup coupling. The intuition is that the renormalization treats
specifically the effect of the main resonances at each scale. Their Fourier modes
 correspond to the rational approximants of ${\u \omega}_0$, which 
produce the small denominators in the perturbation expansion and are responsible
of the eventual breakup of the torus. Our results show that renormalization
flattens out the step structure
 of the jumps in $\varepsilon_c(L)$, and gives a very accurate value for the breakup
coupling, already for a representation with a small number of Fourier coefficients.

The KAM-RG transformation is composed of four steps:
(1) a KAM transformation as described above,
(2) a shift of the resonances: a canonical change of coordinates that maps 
the next pair of  daughter resonances of the sequence of rational approximants
into the two main resonances,
(3) a rescaling of energy (or equivalently of time), and
(4) a rescaling of the action variables (which is
	      a generalized canonical transformation).

The aim of this transformation is to go from one scale to a smaller scale. 
The last three steps are implemented as follows:
The two main resonances 
$(1,0)$ and
$(1,1)$ are replaced by the 
next pair of daughter resonances 
$(2,1)$ and
$(3,2)$. This change is done via a 
canonical transformation 
$(\u{A},\u{\varphi})\mapsto (\mat{N}^{-2}\u{A},\mat{N}^2\u{\varphi})$ with 
$$
\mat{N}^2=\d\left(\begin{array}{cc} 2 & 1 \\ 1 & 1 \end{array}\right).
$$
This linear transformation multiplies ${\u \omega}_0$ by $\goldenmean^{-2}$;
therefore we rescale the energy by a factor $\goldenmean^2$ in order to keep
 the frequency at ${\u \omega}_0$.
Then we perform a rescaling of the action variables: we change the  
Hamiltonian $H$ into 
$$
H'({\u A},{\u \varphi})=
\lambda H\left(\d\frac{{\u A}}{\lambda},{\u \varphi}\right)
$$
with $\lambda$
such that the mean value of $m$ remains equal to $1$.
This magnifies the size of the daughter resonances,
 and places them at the location of the original ones.\\
The functions $m$, $g$, $f$ are rescaled and ${\u \Omega}=(1,\alpha)$ 
is transformed as
 \begin{equation}
  \alpha \mapsto \alpha'=\frac{1+\alpha}{2+\alpha}.\label{eqn:alpha}
\end{equation}
An
effect of the shifting and rescaling is thus  to  select
$\alpha=\goldenmean^{-1}$: ${\u \Omega}$ converges under  successive
iterations  to ${\u \Omega}_*$, which
 is orthogonal to ${\u \omega}_0$ and is the unstable 
eigenvector of $\mat{N}^2$ with the largest eigenvalue $\goldenmean^2$.

In Fig.\ 1 we show the values of the breakup coupling $\varepsilon_c(L)$,
calculated by this KAM-RG transformation. We obtain
$\varepsilon_c\in[0.02758,0.02760]$, which is in very good agreement with the
value  $\varepsilon_c=0.0275856$ obtained with the Greene criterion.

 The improvement with respect  to the  KAM iteration is not only
quantitative; the disappearance of the steps is a strong evidence that the KAM-RG
transformation we have constructed captures the essential physical
mechanism of the breakup of the tori.

\paragraph*{Nontrivial fixed point.---}
By iterating the KAM-RG transformation starting from a point on the breakup
surface, we observe that the process converges to a nontrivial fixed point
$H_*$,
which we characterize by the Fourier coefficients of the three functions
 $f_*,g_*,m_*$ and ${\u \Omega}_*=(1,\goldenmean^{-1})$.
Figure 2 shows the weight of the Fourier coefficients of $f_*$. We observe that
the nonzero coefficients are strongly concentrated on a band around the direction 
$D_\perp$ of the frequency vector ${\u \omega}_0$, i.e., perpendicular to
the line of resonances.
The  decrease of the size of the coefficients along $D_\perp$
is quite slow.
The Fourier coefficients of $g_*$ and $m_*$ have a
similar overall behavior, but they decay faster in the $D_\perp$ direction. 
By linearizing the KAM-RG transformation around $H_*$, we calculate
the critical exponents. 
There is only  one with $|\delta|>1$. This 
implies that the critical surface, which is
the stable manifold 
of $H_*$, is of codimension one.
The value we obtain for the relevant critical exponent is $\delta\in[2.67,2.68]$
which is quite close to the one obtained by MacKay for area-preserving
maps\cite{mackay} ($\delta=2.65$), and to the one obtained by Escande 
{\it et al.} with the approximate scheme ($\delta=2.75$)
\cite{escandeetal}.\\
We also obtain an attractive cycle of period three,
as it had also been encountered in area-preserving maps\cite{greenemao}.
If odd perturbations are included, higher period can appear.
In the Hamiltonian case these cycles are simply related to the
fixed point $H_*$ by symmetry~\cite{mackayproc}.

In conclusion, we have shown that the KAM-RG technique can be implemented numerically with high
accuracy. The results show that the KAM-RG transformation follows closely
the mechanism of the breakup of the invariant tori by a sequence of resonances.
Their effect at all scales leads to the universal self-similar structure of
the critical tori. 
We have presented results for two degrees of freedom, but the extension 
to three\cite{mackaymeissstark} or higher dimensional systems
should be accessible. 
The extension of the scheme to other frequencies that are quadratic irrationals
is relatively clear, but the case of a general irrational frequency
will involve some qualitatively new features.

We thank G. Gallavotti and G. Benfatto who initiated this approach in 1987
\cite{gallben}. Our discussions with them
were very helpful for this project.
We also thank A. Celletti for providing the program 
to determine the critical coupling by Greene's criterion.
Support from the EC contract ERBCHRXCT94-0460 for the project
``Stability and universality in classical mechanics''
and from  the Conseil R\'egional de Bourgogne is acknowledged.

\begin{figure}
\large
\unitlength=1cm
\centerline{
\begin{picture}(15,10)
\put(1.5,0){\psfig{figure=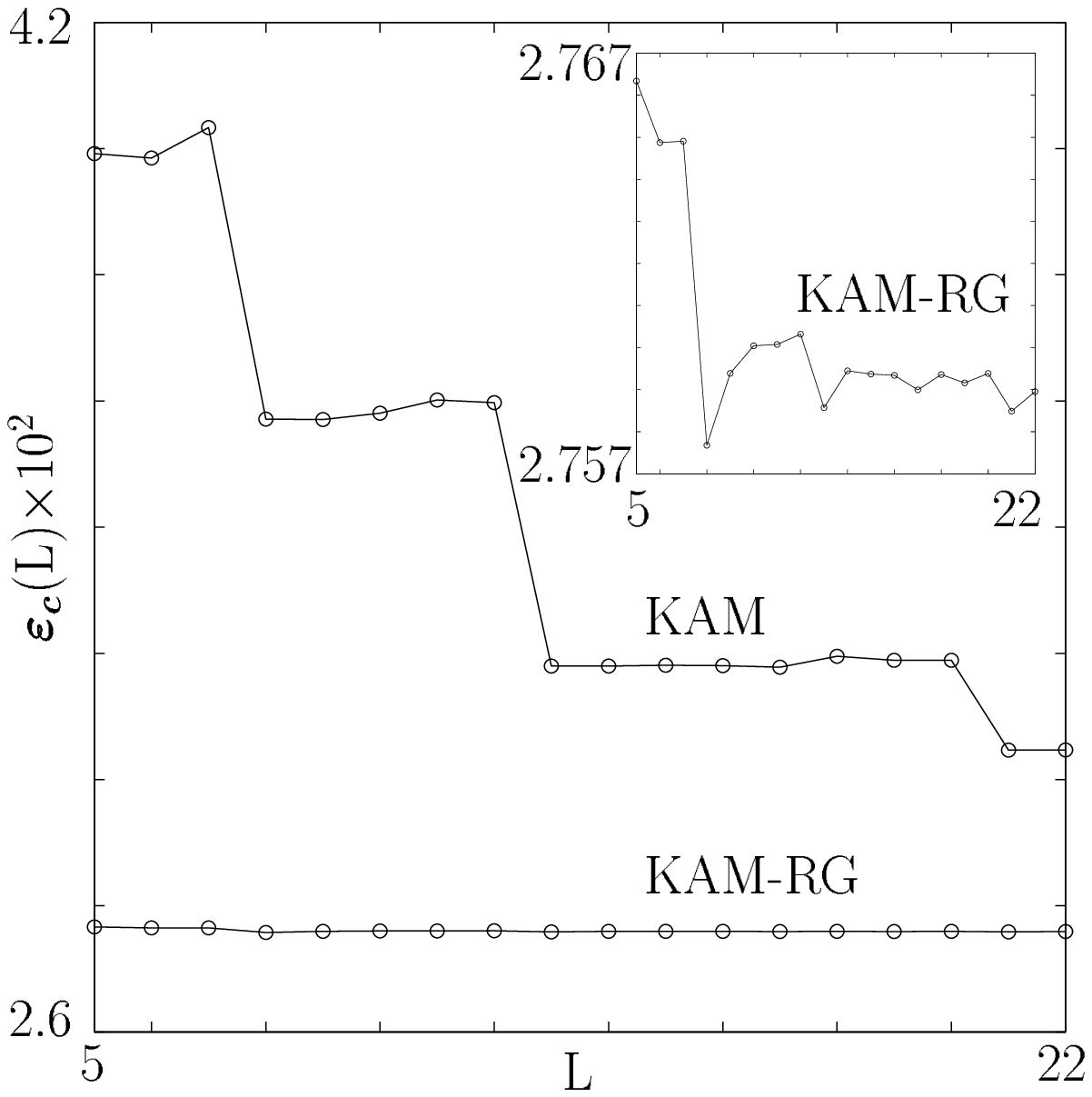,height=15cm,width=10cm}}
\normalsize 
\put(3,2){FIG.\ 1. Critical coupling $\varepsilon_c(L)$ as a function of $L$, the}
\put(3,1.6){size of the cell $\mathcal{C}_L$ containing $(2L+1)^2$ 
Fourier coefficients.}
\put(3,1.2){The upper curve corresponds to the KAM transformation,}
\put(3,0.8){and the lower one
 (enlarged in the inset) to the KAM-RG}
\put(3,0.4){transformation.}
\end{picture}}
\end{figure}

\begin{figure}
\large
\unitlength=1cm
\centerline{
\begin{picture}(15,10)
\put(1.5,0){\psfig{figure=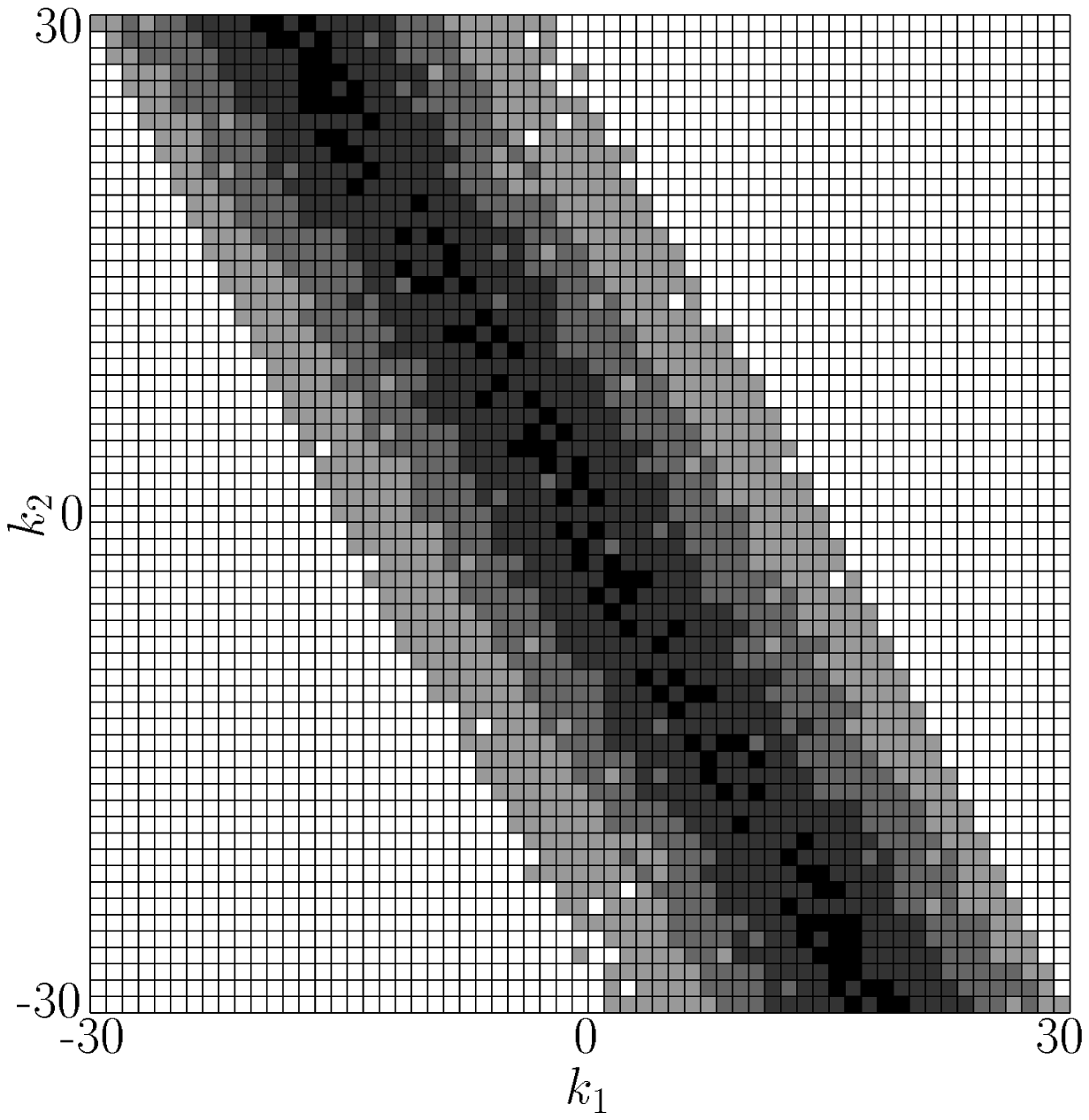,height=15cm,width=10cm}}
\normalsize 
\put(3,2){
FIG.\ 2. Weight of the Fourier coefficients of $f_*$:}
\put(3,1.6){White: $\, <10^{-10}$, grey levels: $[10^{-10},10^{-7}]$,
 $[10^{-7},10^{-5}]$,}
\put(3,1.2){$[10^{-5},10^{-3}]$, black: $[10^{-3},10^{-2}]$.}
\end{picture}}
\end{figure}

\newpage

\end{document}